\documentstyle[aps,prl,twocolumn,tighten,psfig]{revtex}  

\def\issue(#1,#2,#3){{\bf #1}, #2 (#3)} 

\def\opcit(#1){ {\em op. cit.}, #1}

\def\APP(#1,#2,#3){Acta Phys.\ Polon.\ \issue(#1,#2,#3)}
\def\ARNPS(#1,#2,#3){Ann.\ Rev.\ Nucl.\ Part.\ Sci.\ \issue(#1,#2,#3)}
\def\CPC(#1,#2,#3){Comp.\ Phys.\ Comm.\ \issue(#1,#2,#3)}
\def\CIP(#1,#2,#3){Comput.\ Phys.\ \issue(#1,#2,#3)}
\def\CJP(#1,#2,#3){Chin.\ J.\ Phys. (Taipei)\ \issue(#1,#2,#3)}
\def\EPJC(#1,#2,#3){Eur.\ Phys.\ J.\ C\ \issue(#1,#2,#3)}
\def\EPJD(#1,#2,#3){Eur.\ Phys.\ J. Direct\ C\ \issue(#1,#2,#3)}
\def\IEEETNS(#1,#2,#3){IEEE Trans.\ Nucl.\ Sci.\ \issue(#1,#2,#3)}
\def\JHEP(#1,#2,#3){JHEP\ \issue(#1,#2,#3)}
\def\MPL(#1,#2,#3){Mod.\ Phys.\ Lett.\ \issue(#1,#2,#3)}
\def\NP(#1,#2,#3){Nucl.\ Phys.\ \issue(#1,#2,#3)}
\def\NIM(#1,#2,#3){Nucl.\ Instrum.\ Meth.\ \issue(#1,#2,#3)}
\def\PL(#1,#2,#3){Phys.\ Lett.\ \issue(#1,#2,#3)}
\def\PRD(#1,#2,#3){Phys.\ Rev.\ D \issue(#1,#2,#3)}
\def\PRL(#1,#2,#3){Phys.\ Rev.\ Lett.\ \issue(#1,#2,#3)}
\def\SJNP(#1,#2,#3){Sov.\ J. Nucl.\ Phys.\ \issue(#1,#2,#3)}
\def\ZPC(#1,#2,#3){Zeit.\ Phys.\ C \issue(#1,#2,#3)}

\def\be {\begin{equation}}
\def\ee {\end{equation}}
\def\bea {\begin{eqnarray}}
\def\eea {\end{eqnarray}}

\def\bc {\begin{center}}
\def\ec {\end{center}}

\def\lapp{\mathrel{\rlap{\raise.5ex\hbox{$<$}} {\lower.5ex\hbox{$\sim$}}}}
\def\gapp{\mathrel{\rlap{\raise.5ex\hbox{$>$}} {\lower.5ex\hbox{$\sim$}}}}
\def\r {\rightarrow}

\def \MHD{m_{H_2}^2}
\def \CH{{\tilde\chi}^{\pm}}
\def \LSP{\tilde\chi_1^0}

\def \MCH{m_{{\tilde\chi}^{\pm}}}
\def \MCHMIN {\MCH^{min}}
\def \ET{\not\!\!{E_T}}

\begin{document}
\draft
\twocolumn[\hsize\textwidth\columnwidth\hsize\csname
  @twocolumnfalse\endcsname
\vspace{-0.5in}
\title{
New Bounds on Slepton and Wino Masses in Anomaly 
Mediated Supersymmetry Breaking Models
}
\author{Amitava Datta\cite{eml1,leave}}
\address{Department of Physics, Visva-Bharati, Santiniketan - 731 235,
India}
\author{Anirban Kundu\cite{eml2}
and Abhijit Samanta\cite{eml3}}
\address{Department of Physics, Jadavpur University,
Calcutta - 700 032, India}
\date{\today}

\maketitle

\begin{abstract}
We show how the spectrum of the minimal anomaly-mediated supersymmetry
breaking model can be constrained from the condition that the electroweak
symmetry breaking
minimum of the scalar potential is the deepest point in the field space. 
Applying the current experimental bounds and scanning over the whole 
parameter space, we rule out selectrons below 378 GeV and staus below 269 GeV,
the numbers having a modest uncertainty.
We also find a new upper bound on the wino-like
chargino mass for a given slepton mass.
This rules out the possibility of slepton pair production at ongoing or 
upcoming colliders like the Tevatron or  the Next Linear Collider
at $\sqrt{s}=500$ GeV, where
pair production of charginos may be the only available signal. 
\end{abstract}
\pacs{PACS number: 12.60.Jv, 14.80.Ly, 14.80.Cp}
] 

\narrowtext

Supersymmetry (SUSY), if it exists, must be broken, and this breaking cannot
take place in the observable sector (OS). Thus one envisages a hidden
sector (HS), whose fields are all singlets under the SM gauge group,
where SUSY is broken. The key question is how to convey the breaking to the
OS. One option is to consider a contact interaction between the HS and the
OS fields in the Kahler potential, suppressed by the Planck mass squared.
This tree-level interaction induces SUSY breaking in the OS; such models
are generically known as supergravity (SUGRA) type models, where the
gravitino mass is of the order of 1 TeV.

Recently, one came to note that if the OS and the
HS live in two distinct 3-branes separated by a finite distance along a
fifth compactified dimension, there is no tree-level term in the Kahler
potential that transmit SUSY breaking from the HS to the OS. 
However, a superconformal anomaly may induce the SUSY breaking in
the OS (this term is present in the SUGRA type models too,
but is suppressed in comparison to the usual soft-breaking terms).
To generate the weak scale masses of the
sparticles, the gravitino mass must be of the order of tens of TeV. Such
models are generically known as anomaly-mediated SUSY breaking (AMSB)
models \cite{amsb,amsb2}.

AMSB, alongwith the radiative electroweak symmetry breaking condition,
should fix the sparticle spectrum completely in terms of three parameters:
$m_{3/2}$ (the mass of the fermionic component of the compensator
superfield,
and equal to the gravitino mass), $\tan\beta$ (ratio of the vacuum
expectation values (VEV) of the two Higgs fields), and $sign(\mu)$.
The gaugino masses $M_1$, $M_2$ and $M_3$, and the trilinear
couplings (generically denoted by $A$) can be obtained from the
relevant renormalization group (RG)
$\beta$-functions and anomalous dimensions. The sfermion masses, as well
as the Higgs mass parameters, are also determined by $m_{3/2}$;
unfortunately,
for sfermions that do not couple to asymptotically free gauge groups ({\em
i.e.}, both right and left sleptons), the masses come out to be tachyonic.
The remedy is sought by putting a positive definite mass squared term
$m_0^2$
in the GUT scale boundary conditions. This is not exactly an {\em ad hoc}
prescription; there are a number of physical
motivations for the introduction of such a term, mostly related to the
presence of extra field(s) in the bulk. Such models with a universal $m_0$
for all scalars are called the minimal AMSB (mAMSB) models
\cite{amsb,amsb2}.
The phenomenology of such models has been at the focus of attention of
many recent works \cite{ggw,feng,amsb-pheno,probir,baer1,baer2},
and we also confine our discussions within the scope of mAMSB models.

With four free parameters in the model, one can determine the complete
particle spectrum. A few key observations can be immediately made
\cite{amsb,amsb2,ggw,feng,amsb-pheno}: \\
(i) The lighter chargino $\CH$ is almost degenerate with the lightest 
neutralino $\LSP$, which we assume to be the lightest supersymmetric 
particle (LSP). Both of them are heavily dominated by the wino component. 
The near degeneracy  leads to
the most striking experimental signature of AMSB models, based on
the ``nearly invisible'' decay of the relatively long lived
$\CH$ to the LSP and a soft charged pion \cite{chen}.  A heavily ionizing
charged track in the vertex detector ending in a soft pion is taken to be
the smoking gun of such models.\\
(ii) $M_1$, $M_2$ and $M_3$ increase with $m_{3/2}$ but are insensitive
to $m_0$ for all practical purpose. \\
(iii) Sfermion masses increase linearly with $m_0$, but also depend on
the precise value of $m_{3/2}$. \\
(iv) Right and left sleptons of the first two generations
are highly degenerate. Stau is the lightest slepton.

From these observations one can understand the currently available bounds
on the mAMSB spectrum and the correlations among them. They are summarized
in the following (see Figure 1):\\
(a) There is a lower limit on $m_{3/2}$ coming from the  lower bound 
$\MCHMIN = 86$ GeV \cite{chargino} on charginos decaying through the soft 
pion mode from  direct searches at LEP.
This limit depends on a number of uncertainties to be elaborated below,
but is $\sim 28$-32 TeV. Region III of Fig.\ 1 is ruled out by this
constraint.\\
(b) For a given $m_{3/2}$, there is a lower bound
on $m_0$ below which either lighter stau($\tilde\tau_1$) is the LSP 
or sleptons would have
been observable at present-day colliders. This is shown as the boundary 
between regions II and IV in Fig.\ 1.\\ 
(c) For a very limited region of the parameter space with 
$\tan\beta \approx 3$ and relatively small $m_{3/2}$, a further region
is ruled out \cite{baer2} from the bound on neutral Higgs mass 
from LEP (106 GeV for small $\tan\beta$ and 88 GeV for large $\tan\beta$) 
\cite{higgs}. The limit  disappears for larger  $m_{3/2}$ and/or 
large $\tan\beta$. Thus over a significant region of the currently available
allowed parameter space (APS), sleptons as light as 150 GeV cannot be
ruled out.

The main result of this letter is that over the whole APS there is a rather
stringent lower bound on the slepton mass with important bearings on the search
prospects at present or future colliders. An interesting upper bound on
the wino mass also emerges. The main ingredient  of our
analysis is the so-called ``unbounded from below'' (UFB) directions of the
scalar potential.

The real minimum of the SUSY scalar potential
occurs along the direction where only the neutral components of two Higgs
fields acquire nonzero VEVs. However, with a number of other scalar fields in
the theory carrying charge and/or color, it may be possible to find some 
other direction(s) where
nonzero VEVs to the corresponding scalar fields make the potential deeper
than the real minimum, or even unbounded from below.

It is well known that the determination of these dangerous direction
from the tree-level potential alone may not be reliable \cite{gamberini},
and at least one-loop corrections should be included to get approximately
reliable results. The addition of such terms on the other hand makes the
minimization program rather involved and practically undoable when a large
region of the parameter space has to be scanned. As a compromise the
tree-level potential is analyzed  at an optimum
mass scale where the one-loop corrections are estimated to be small.
If for a
particular choice of the SUSY parameters, one gets even one such UFB
direction, that set is not allowed.

UFB directions were classified in broadly three categories in a model
independent way  by Casas {\em et al} \cite{casas}.
They are chosen in such a way that the F terms are zero and the
D terms either vanish or are kept under control. These directions, known
as UFB-1,2 and 3 (eqs.\ (20), (26) and (33) of \cite{casas}),
are characterized by nonzero VEVs
of the Higgs fields as well as of various slepton and/or squark fields.

Casas {\em et al} illustrated  the constraints emerging from various UFB
conditions numerically
in the context of SUGRA models with a common scalar mass ($m_0$) and 
low values of $\tan\beta$ only.
This work was supplemented by considering large values of $\tan\beta$
motivated by possible $b$-$\tau$ and $t$-$b$-$\tau$ Yukawa unification
scenarios in context of SO(10) GUTs \cite{abhijit}.
However, in such models, the common trilinear
coupling at the GUT scale, $A_0$, is a free parameter, and the applicability
of the UFB conditions depend crucially on the value of $A_0$. For example,
they are quite restrictive for negative $A_0$ but loses their constraining
power for $A_0 > 0$ \cite{abhijit}.

It is precisely here that the AMSB models score over the SUGRA ones. For,
in AMSB models, $A$-terms at the GUT scale (they do not
unify, by the way) are not free parameters but are completely determined
by the other four free parameters of the theory. Thus, the constraints that
one may obtain on the APS are independent of one major source of uncertainty
in the SUGRA models.

In this letter we find strong  limits on $m_0$ for various values of
$m_{3/2}$, $\tan\beta$ and $sign(\mu)$. The physical spectrum as well as
the values of the running masses at different energy scales are determined
with the ISAJET 7.51 code \cite{baer1}.
We demand the
tree-level minima of the  potential along an UFB direction
(evaluated at a properly chosen scale $Q_c$ as discussed above),
to be always shallower than the true minimum  evaluated
at a scale $Q_r=\sqrt{m_{\tilde t_L}m_{\tilde t_R}}$.
In presenting our results (fig.\ 1 and the tables) we follow the prescription
laid down by Casas {\em et al} in choosing the scale $Q_c$ (eq.\ 33 of
\cite{casas}). This
prescription, however, gives an order of magnitude estimate of $Q_c$ rather
than a precise value. Therefore, to check the
robustness of our limits, we enforce the condition $V_{UFB}(Q_{UFB})
< V_{realmin}(Q_r)$ for $Q_{UFB}$ varying from $10Q_c$ to $Q_c / 10 $.
This, we think, is rather conservative and shall quote below the 
possible relaxation of our limits. 

The allowed and ruled out regions of the parameter space are shown in Fig.\ 1 
for $\tan\beta=5$ and the choice of $Q_c$ as
stated above.  $\mu$ is taken to be negative,
but the bounds are not very sensitive on the sign of $\mu$ (see
below). The main
characteristics that follow from the figure are as follows:

\noindent (i) 
There is a lower bound on $m_0$ for a particular value of $m_{3/2}$, and
this bound increases almost linearly with increasing $m_{3/2}$.
This is because for low $m_0$, the sum of the Higgs mass parameter squared
$\MHD$ and the left-handed slepton mass squared $m_{L_i}^2$ (see eq.\ (33) of
\cite{casas}) becomes
so negative as to violate the UFB-3 constraint. Thus, there exists a minimum
value of $m_0$ (indicated by the boundary between regions I and II in
Fig.\ 1) above which the real minimum of the scalar potential is
the deepest point in the field space. This strong lower bound comes for
$m_{3/2}$ corresponding to $\MCHMIN$. Translated into slepton masses, the
bound reads $m_{\tilde e} \geq 425$ GeV and $m_{\tilde\tau_1} \geq 416$ GeV.

\noindent (ii) For a given $m_{3/2}$, 
larger $\tan\beta$ leads to stronger lower bounds on $m_0$ (see Table 1). 
This is due to the fact that the stau becomes lighter as $\tan\beta$ increases
which makes the UFB-3 condition more restrictive.

\noindent (iii) 
For a given slepton mass, there exists both lower and upper limits
on the chargino mass. The lower limit comes from $\MCHMIN$.
We find a strong upper limit too
on $\MCH$; above this, the UFB conditions are violated (see Table 2).

\noindent (iv)
The UFB-3 constraint, by itself, ensures a neutralino LSP; one need 
not separately impose the condition that slepton/sneutrino should not be 
the LSP. In other words, region IV of Fig.\ 1 is actually a subset of 
region II.

The variations in the limit on $m_0$, for various choices of
$sign(\mu)$ and $\tan\beta$, are shown in Table 1. 
However, these bounds depend crucially
on the experimental bound on the mass of this type of chargino and will
get much stronger if this bound improves.

A theoretical uncertainty in our limits arises due to the following
reason. The running chargino mass at a given scale is related to
$m_{3/2}$. When this mass is compared with the chargino mass bound
it should be translated to the corresponding pole mass which is related
to the running mass via certain weak threshold corrections \cite{ggw}
(see in particular eqs.\ (16) and (18)). These
corrections are not incorporated in ISAJET 7.51.
On the other hand the corrections are  non-negligible. After
including this correction our strongest bound (Table 1) on $m_{\tilde e}$
is relaxed to 361 GeV. We have checked that for other points in the APS this
relaxation amounts to 20-25 GeV at the most for $m_{\tilde e}$ and $m_{\tilde
\tau_1}$.

The uncertainty in the scale $Q_{UFB}$ as discussed above may also relax
our limits. For example, at a scale $10Q_c$ our strongest bounds (Table
1, set marked by an asterisk) on 
slepton masses are relaxed by about 25 GeV. Using both the weak threshold
corrections and the modified scale we obtain $m_{\tilde e}\geq 335$ GeV.
This, we believe, is the {\em most 
conservative} statement where allowance has been made for all possible
conspiracies to weaken the bound.

In Table 2 we have similarly illustrated the variations in the upper
limit on $\MCH$ for given values of $m_0$. Inclusion of weak threshold 
corrections, however, modestly weakens this bound ($\sim 20\%$ \cite{ggw}). 
The only other
upper bound available in the literature is $M_2 < 200$ GeV, which comes
from naturalness arguments \cite{feng}. We wish to stress that the 
requirement of no finetuning, though intuitively appealing, is difficult to
quantify. It is gratifying to note that we are getting similar or even stronger
limits from a well defined physical principle.

In the mAMSB model  $\CH$ decays dominantly into a nearly degenerate
$\LSP$ and a soft $\pi$ after travelling typically through a distance of
a few cms \cite{amsb-pheno} and traverses several layers of the vertex 
detector in the process. 
The signature of this $\CH$ and/or the soft $\pi$ in the detector 
during the analysis stage may help to reduce the background to a
negligible level. However, other particles in the final state in addition
to the $\CH$s are needed to trigger the event. Various suggestions and
their viability in the light of our bounds are summarized below.

In \cite{ggw,probir} the proposed signals were based on the decays
$\tilde e_L\r e + \LSP, \nu + \CH$ and
$\tilde \nu \r \nu+\LSP, e+\CH$.
Thus a $\tilde\nu$-$\tilde\ell_L$ pair produced at hadron colliders,
{\em e.g.}, would lead
to a $2\ell+ \CH + \LSP$ where the hard leptons  can be triggered on.
It was estimated in \cite{ggw} that a detectable signal can be found
at the Tevatron for $m_{\tilde\nu} < 200$ GeV. In view of the new bounds such
signals are strongly disfavoured. A similar signal from $\tilde\ell_L$ pairs
produced at the Next Linear Collider (NLC) \cite{probir}
is also improbable at the early versions of the NLC, say at $\sqrt{s} = 500$
GeV. Thus a properly constrained
mAMSB model leaves open the possibility of the Large Hadronic Collider
(LHC) as the only source of slepton signals in the near future.

On the other hand, our upper bounds on $\MCH$ as a function of $m_0$
reveal that the chargino will necessarily be in the striking range of the
Tevatron for a wide range of
slepton masses. The signal from chargino pair plus mono-jet, where the jet is
essentially required for triggering, was recommended 
by Feng {\em et al} in \cite{amsb-pheno}. It was
further pointed out that typical discovery reaches are 
$\MCH=$ 140, 210, 240 GeV for $c\tau=3, 10, 30$ cm, where $\tau$ is the 
mean lifetime of $\CH$. Our upper bounds
on $\MCH$ reinforces the prospect of this channel.

The signal $e^+e^-\r \tilde\chi^+\tilde\chi^-\gamma$ where the hard photon 
triggers the event \cite{chen}, has also been proposed. Our bounds 
also strenghthens the possibility of 
observing this signal at an early version of the NLC. Moreover,
if $\MCH$ is not too close to the kinematic limit, one can determine 
$\MCH$ from the distribution of $(p_{e^-} + p_{e^+} - p_{\gamma})^2
> 4 \MCH^2$. In addition, if the $\tilde\ell_L$ is indeed light (say, 400-500 GeV),
which is the case in mAMSB over a large region of the APS, some
useful hints on $m_{\tilde\ell_L}$ may come as a bonus, from the size of the 
cross-section. Due to the destructive interference between the s-channel 
$\gamma,Z$ mediated diagrams and the t-channel $\tilde\nu$ exchange diagram
the cross-section will be smaller for lighter $\tilde\ell_L$ \cite{shyama}.
Competing models with nearly invisible $\CH$ like the string motivated
model ($m_{\tilde\ell_L}\approx 1$ TeV) 
of \cite{chen} would yield much larger cross sections for comparable $\CH$
mass. Models with nearly degenerate Higgsino dominated chargino and
LSP, on the other hand, can be distinguished from the AMSB model by measuring
the cross-section with right polarized electrons.

The region of the $m_0-m_{3/2}$ plane that can be probed at the LHC 
via the conventional $m$ lepton + $n$ jets + $\ET$ signals has been given
in \cite{baer2}. A significant part of this region is, however, ruled out 
by the new bounds presented in this paper.

One could have evaded our bounds by assuming the 
universe to be in a standard model like false vacuum, separated from a
charge and color breaking
true vacuum by a barrier with a tunnelling time too large compared
to the age of the universe \cite{hall}. 
The calculation, which is relatively straightforward for a single scalar field,
is rather complicated and uncertain in SUSY models. Moreover, it cannot be 
checked against experimental data. 
Thus the false vacuum option is at best a theoretically interesting 
alternative to the much simpler true vacuum hypothesis
favored by the principle of Occam's razor. 
We, however, wish to stress that even if the alternative scenario happens
to be the correct one our bounds do not lose their significance. If,
{\em e.g.}, one discovers the
AMSB spectrum (almost degenerate lowest-lying gauginos, maximal mixing between
right-and left-handed smuons etc.) alongwith sleptons violating our bounds,
it may be the first indication that we are living in a false vacuum. This,
we note, is experimentally a much simpler way to find out our 
precarious existence in a false vacuum, compared to what 
happens in the minimal SUGRA models, where, {\em e.g.},
one has to determine the trilinear $A$ parameter precisely to get the same
information. 

{\em Acknowledgements}:
The work of AD was supported by DST, India (Project No.\ SP/S2/k01/97)
and BRNS, India (Project No.\ 37/4/97 - R \& D II/474). AK was supported
by BRNS, India (Project No.\ 2000/37/10/BRNS).
AS acknowledges CSIR, India, for a research fellowship.

\begin{figure}[htb]
\centerline{
\psfig{file=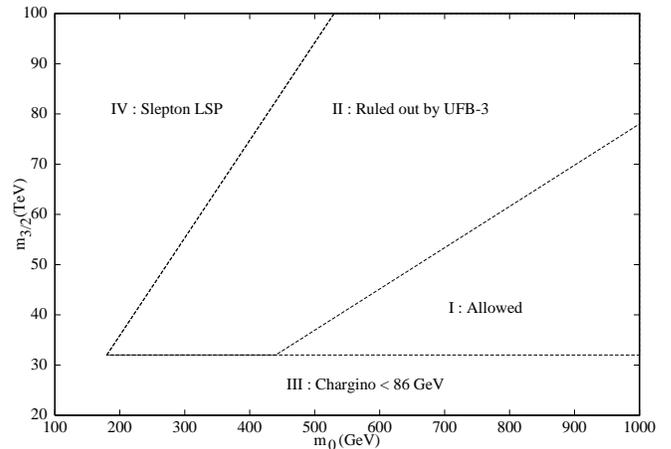,width=\linewidth,angle=270}
}
\caption{Allowed parameter space for the mAMSB model for $\tan\beta=5$ and 
$\mu < 0$. Already existing constraints are depicted as regions III and
IV, while our constraints fall in region II. Only region I remains allowed.}
    \label{fig:figure1}
\end{figure}
\begin{table}[htbp]
\caption{\sl{Lower bounds on $m_0$ and corresponding slepton masses for
different sets of input parameters (see text). The limit $\MCHMIN
=86$ GeV has been used. The set marked by the asterisk corresponds to the 
minimum $m_{\tilde e}$ consistent with $\MCHMIN$ and the Higgs mass limit
as discussed in the text. The set marked by the dagger corresponds to
the minimum value of $m_{\tilde\tau_1}$.}}
\begin{center}
\begin{tabular}{||c|c|c|c|c|c||}
$\tan\beta$ & sign($\mu$) & $m_{3/2}$ & $m_0$ & 
$m_{\tilde e}$ & $m_{\tilde\tau_1}$ \\
& & (TeV) & (GeV) & (GeV) & (GeV)\\
\hline
3.3 ($\ast$) & + & 27.9 & 395 & 378 & 372\\
5 & + & 28.6 & 406 & 389 & 380 \\
5 & $-$ & 31.6 & 446 & 425 & 416\\
35 & + & 30.2 & 457 & 440 & 350 \\
35 & $-$ & 30.8 & 462 & 444 & 355\\
59 ($\dag$) & + & 30.4 & 484 & 468 & 269\\
\end{tabular}
\end{center}
\end{table}
\begin{table}[htbp]
\caption{\sl{Upper bounds on $m_{3/2}$ and corresponding $\MCH$ for
given values of $m_0$. Weak threshold corrections
(see text) weaken the bound by about 20\%.}}
\begin{center}
\begin{tabular}{||c|c|c|c|c||}
$\tan\beta$ & sign($\mu$) & $m_0$ (GeV) & $m_{3/2}$ (TeV) & $\MCH$ (GeV)\\
\hline
5 & + ($-$) & 500 & 35.8 (35.7) & 106 (98)\\
5 & + ($-$) & 700 & 51.6 (51.5) & 150 (144)\\
5 & + ($-$) & 1000 & 76.0 (75.9) & 216 (212)\\
35 & + ($-$) & 500 & 33.3 (33.5) & 95 (94)\\
35 & + ($-$) & 700 & 48.2 (48.5) & 137 (136)\\
35 & + ($-$) & 1000 & 71.1 (71.5) & 200 (200)\\
\end{tabular}
\end{center}
\end{table}

\end{document}